\documentclass[lettersize,journal]{IEEEtran}
\usepackage{amsmath,amsfonts}
\usepackage{algorithmic}
\usepackage{algorithm}
\usepackage{array}
\usepackage[caption=false,font=normalsize,labelfont=sf,textfont=sf]{subfig}
\usepackage{textcomp}
\usepackage{stfloats}
\usepackage{url}
\usepackage{verbatim}
\usepackage{graphicx}
\usepackage{cite}
\usepackage{amsmath}
\usepackage{amsfonts}
\usepackage{tikz}
\usepackage{graphicx}

\usepackage{amsmath,amsfonts}
\usepackage{algorithmic}
\usepackage{array}
\usepackage[caption=false,font=normalsize,labelfont=sf,textfont=sf]{subfig}
\usepackage{textcomp}
\usepackage{stfloats}
\usepackage{url}
\usepackage{verbatim}
\usepackage{graphicx}
\def\BibTeX{{\rm B\kern-.05em{\sc i\kern-.025em b}\kern-.08em
    T\kern-.1667em\lower.7ex\hbox{E}\kern-.125emX}}
\usepackage{balance}

\usepackage{amsmath}
\usepackage{amsfonts}
\usepackage{tikz}
\usepackage{graphicx}
\usepackage{graphicx}
\usepackage{framed}
\usepackage[normalem]{ulem}
\usepackage{amsmath}
\usepackage{amsthm}
\usepackage{amssymb}
\usepackage{amsfonts}
\usepackage{enumerate}
\usepackage[bb=boondox]{mathalfa}
\usepackage{lipsum}  
\usepackage{multirow}
\usepackage{booktabs}

\newtheorem*{remark*}{Remark}

\usepackage{tikz}
\usetikzlibrary{shapes}
\newcommand{\minimize}[1]{\underset{{#1}}{\text{minimize}}}

\DeclareMathSymbol{\shortminus}{\mathbin}{AMSa}{"39}

\newcommand{\st}{\text{subject to}}

\usepackage{mathtools}
\newcommand\norm[1]{\lVert#1\rVert}

\usepackage{esvect}

\definecolor{maincolor}{HTML}{032F99}
\definecolor{blue}{RGB}{0,0,0}
\definecolor{red}{HTML}{e05a87}
\definecolor{green}{HTML}{228b22}

\usepackage{nomencl}
\makenomenclature
\usepackage{etoolbox}
\renewcommand\nomgroup[1]{%
  \item[\bfseries
  \ifstrequal{#1}{A}{\it Problem dimension}{%
  \ifstrequal{#1}{B}{\it Optimization parameters}{%
  \ifstrequal{#1}{C}{\it Optimization variables}{}}}%
]}

\begin{document}
\begingroup
\allowdisplaybreaks

\title{Agent \underline{Co}ordinatio\underline{n} via \underline{C}ontext\underline{u}al \underline{R}egression (AgentCONCUR) for Data Center Flexibility}

\author{
\IEEEauthorblockN{Vladimir Dvorkin, {\it Member, IEEE}}\\
\thanks{The author is with the Department of Electrical Engineering and Computer Science at the University of Michigan, Ann Arbor, MI,  USA. 
}
}


\maketitle

\begin{abstract}
A network of spatially distributed data centers can provide operational flexibility to power systems by shifting computing tasks among electrically remote locations. However, harnessing this flexibility in real-time through the standard optimization techniques is challenged by the need for sensitive operational datasets and substantial computational resources. To alleviate the data and computational requirements, this paper introduces a coordination mechanism based on contextual regression. This mechanism, abbreviated as AgentCONCUR, associates cost-optimal task shifts with public and trusted contextual data (e.g., real-time prices) and uses regression on this data as a coordination policy. Notably, regression-based coordination does not learn the optimal coordination actions from a labeled dataset. Instead, it exploits the optimization structure of the coordination problem to ensure feasible and cost-effective actions. A NYISO-based study reveals large coordination gains and the optimal features for the successful regression-based coordination.
\end{abstract}


\section*{Nomenclature}
The main symbols used in this paper are stated below. Additional symbols are
defined in the paper where needed.

\subsection{Dimensions}
\begin{IEEEdescription}[\IEEEusemathlabelsep\IEEEsetlabelwidth{x}]
\item[$b$] Number of power system buses
\item[$l$] Number of transmission lines
\item[$m$] Number of data center users
\item[$n$] Number of data centers forming a network
\item[$k$] Number of virtual links connecting data centers
\item[$q$] Length of a training dataset
\textcolor{blue}{\item[$t$] Length of the day-ahead planning horizon}
\end{IEEEdescription}
\subsection{Optimization Parameters}
\begin{IEEEdescription}[\IEEEusemathlabelsep\IEEEsetlabelwidth{xx}]
\item[$A$] Incidence matrix of the network of data centers 
\item[$c$] Vector of the $1^{\text{st}}-$order cost coefficients
\item[$C$] Matrix of the $2^{\text{nd}}-$order cost coefficients
\item[$\textcolor{blue}{c^{\text{on}}}$] \textcolor{blue}{Vector of start-up cost coefficients}
\item[$d$] Vector of the electric power loads
\item[$F$] Matrix of the power transfer distribution factors
\item[$\overline{f}$] Vector of the maximum power transmission capacity
\item[$G$] Matrix of distances between users and data centers
\item[$\overline{p}$] Vector of the maximum generation capacities
\item[$\underline{p}$] Vector of the minimum generation capacities
\textcolor{blue}{\item[$\overline{p}^{\uparrow/\uparrow}$] Vectors of maximum ramp up/down capacity
\item[$\overline{p}_{\text{up}}$] Vectors of maximum ramp up capacity at start up
\item[$\overline{p}_{\text{dw}}$] Vectors of maximum ramp down capacity at shot down}
\item[$s$] Vector of the linear load shedding cost coefficients
\item[$x$] Vector of contextual features 

\item[$\alpha$] Latency loss parameters (\% of the nominal latency)
\item[$\Gamma$] Matrix converting computing loads to electric loads
\item[$\delta$] Vector of user computing demands 
\item[$\varepsilon$] Regularization parameter for feature selection 
\item[$\varrho$] Regularization parameter for task allocation 
\item[$\omega$] Vector of renewable power injections
\end{IEEEdescription}

\subsection{Optimization Variables}
\begin{IEEEdescription}[\IEEEusemathlabelsep\IEEEsetlabelwidth{x}]
\item[$\textcolor{blue}{c^{\text{up}}}$] \textcolor{blue}{Vector of auxiliary variables to model start-up}
\item[$\ell$] Vector of electric power load shedding  
\item[$p$] Vector of generation dispatch decisions 
\item[$\textcolor{blue}{u}$] \textcolor{blue}{Vector of unit commitment decisions}
\item[$W$] Matrix allocating user demand among data centers 
\item[$\beta_{0}$] Vector of regression intercept coefficient   
\item[$\beta^{\circ}$] Matrix of regression coefficients associate w/ feature $\circ$   
\item[$\beta$] Vector stacking $\beta_{0}$ and vectorized matrices $\beta^{\circ}$
\textcolor{blue}{\item[$\tilde{\delta}$] Vector of time-shifted user computing demands} 
\item[$\vartheta$] Vector of nominal computing loads of data centers  
\textcolor{blue}{\item[$\tilde{\vartheta}$] Vector of time-shifted computing loads of data centers} 
\item[$\varphi$] Vector of computing task shifts in a data center network
\end{IEEEdescription}\vspace{-0.1cm}

\subsection{Notation} 
Lower- and upper-case letters denote vectors and matrices, respectively. For some matrix $A$, $a_{ij}$ denotes its element at position $(i,j)$. Symbol $^{\top}$ stands for transposition, and $\dot{x}$ denotes the optimal value of $x$. Vectors $\mathbb{0}$ and $\mathbb{1}$ are of zeros and ones, respectively. Operator $\langle \cdot\rangle_{\text{F}}$ is the Frobenius inner product, and $\norm{\cdot}_{p}$ denotes the $p-$norm.

\section{Introduction}
\IEEEPARstart{C}{oordinated} operations of bulk power systems and coupled infrastructures allow for leveraging their complementarity and offsetting operational and economic inefficiencies, thus leading to enhanced performance. Coordination schemes have been proposed to synchronize grid operations with power distribution \cite{papavasiliou2018coordination}, natural gas \cite{ratha2020affine}, water \cite{stuhlmacher2020water}, and district heating \cite{mitridati2018power} systems, and more recently, a large coordination potential has emerged from the networks of spatially distributed data centers (NetDC) \cite{wierman2014opportunities}. The unique coordination advantage of such networks is in {\it spatial flexibility}, which distributed data centers provide by shifting computing tasks among electrically remote locations. This flexibility resource will be important for future power grids, as electricity demand of data centers is rapidly growing, and is expected to reach 35 GW by 2030 in the U.S. alone \cite{McKisey2023}. Even at present, the coordination potential is significant: training a single GPT-3 language model -- the core of the popular {ChatGPT} chatbot -- consumes as much as 1.3 GWh \cite{patterson2021carbon}. Allocating such energy-intensive tasks in the NetDC is thus likely to predetermine the dispatch cost and emission intensity in adjacent grids.

The growing environmental footprint of computing has encouraged large internet companies to optimize NetDC operations in a carbon-aware manner. Using online emission monitoring tools, such as \textsc{\small\url{WattTime.org}} and \textsc{\small\url{ElectricityMaps.com}}, sthey smartly time and allocate computing tasks in regions with the least emission footprint \cite{james2019low,radovanovic2022carbon}. However, the sole reliance on limited emission data is the form of {\it grid-agnostic} coordination, which respects NetDC constraints yet ignores those of power grids. For {\it grid-aware} coordination, the literature offers three coordination mechanisms: demand response \cite{wierman2014opportunities}, enhanced market design \cite{ruddy2014global,fridgen2017shifting,zhang2020flexibility}, and co-optimization of grid and NetDC operations \cite{kim2016data}. \textcolor{blue}{For the former, the work in~\cite{zhang2022exploring} develops a comprehensive data center management protocol which accounts for a broader range of signals from the power grid, e.g., including not only carbon intensity but also electricity prices and loads; a closely related work also includes \cite{paul2015demand,fang2014using,parolini2011cyber,wang2016renewable}. While demand response protocols follow power grid signals, the benefits from this type of coordination are still limited, as power grid and NetDC operations are still optimized separately.} \textcolor{blue}{The second set of mechanisms integrate the flexibility of data centers within market-clearing algorithms. Although they features robust market properties guaranteeing theoretical efficiency of system operations \cite{werner2021pricing,zhang2022remunerating}},  it remains challenging to fully represent complex data center objectives (e.g., quality of service) and constraints (e.g., latency) via single utility function. \textcolor{blue}{The third set of mechanism propose to co-optimize the grid and NetDC operations within operational planning routines \cite{kim2016data}}, with the full representation of operational constraints, akin to coordination models for conventional energy infrastructures \cite{papavasiliou2018coordination,ratha2020affine,stuhlmacher2020water,mitridati2018power}.  \textcolor{blue}{While this approach suits the day-ahead planning, compensating for any deviation from the planning parameters in real time by resolving this optimization is challenging due to large data and computational requirements.} 

This paper develops a new, regression-based mechanism for grid-aware coordination of power systems and NetDC, termed as AgentCONCUR. The grid-awareness is achieved by embedding the decision-making of NetDC within the OPF optimization of a system operator, and jointly optimizing the objectives of the two systems within one optimization problem. However, unlike optimization-based coordination, AgentCONCUR shifts all computational burden to the planning stage, where a linear regression is trained to act on many operational scenarios. The coordination in the real time is then carried out by a trained regression, which  solely acts on available contextual grid information, such as electricity prices. This way, AgentCONCUR resembles industry practices in \cite{james2019low} and \cite{radovanovic2022carbon} by relying on limited grid data, while also leveraging the optimization structure of power system's and NetDC's problems, thereby ensuring feasibility and economic efficiency of the Power--NetDC coordination. More specifically, this paper provides the following technical contributions:
\begin{enumerate}
    \item To estimate economic potential of \textcolor{blue}{the Power-NetDC coordination} under grid and communication constraints, the paper develops a bilevel co-optimization where power system decision-making is constrained by that of NetDC. Similar to grid-aware models in \cite{zhang2020flexibility}--\cite{zhang2022remunerating}, this model takes the power system perspective, but it represents the NetDC via an embedded optimization problem with customer-oriented objectives and constraints, i.e., minimizing communication latency. \textcolor{blue}{This co-optimization provides the ideal coordination solution, suitable for the day-ahead operational planning  of the two systems.} 

    \item To enable fast and data-light implementation of such coordination \textcolor{blue}{in real-time}, the paper devices a contextual regression policy that approximates the ideal coordination. The policy feasibility and cost-consistency is ensured by the new training optimization which inherits the bilevel optimization structure. Using sufficiently many operational scenarios in training allows for robust and cost-consistent performance across testing scenarios. Furthermore, the proposed training allows for the optimal coordination feature selection, such that the coordination can be made possible at different data requirements. 
\end{enumerate}

\textcolor{blue}{To demonstrate the cost-saving potential of the Power--NetDC coordination, we present a case study based on the system of the New York Independent System Operator (ISO). First, we demonstrate the performance of the day-ahead planning, where unit commitment and dispatch decisions are optimized by leveraging the space-time flexibility of NetDC. We then present the results of AgentCONCUR for peak-hour coordination by means of spatial computing task redistribution.} The results reveal practical trade-offs between the amount of contextual information (features) and the efficiency of coordination. Interestingly, the case study found that the economic coordination between the two systems can be achieved by relying on a single contextual feature, i.e., the electric power load of the largest electricity demand center -- New York City. However, there are more substantial economic gains that come with more data requirements. 

The AgentCONCUR mechanism contributes to the area of optimization with contextual information \cite{munoz2022bilevel}, also referred to as smart predict-then-optimize \cite{elmachtoub2022smart} or decision-focused learning \cite{mandi2023decision}. While most of the prior work in power systems has focused  on decision-optimal \textit{data} predictions, e.g., demand \cite{munoz2022bilevel} or wind power \cite{dvorkin2023price}, AgentCONCUR enables contextual predictions of the optimal coordination \textit{decisions} instead. The mechanism also relates to learning-based grid coordination approaches, such as the one developed for transmission-distribution system coordination in \cite{morales2021learning}. However, the distinct features of the AgentCONCUR mechanism include its small computational requirements at the time of coordination, ability to preserve operational feasibility for coordinating agents, and the optimal coordination feature selection. 

In the remainder, Section \ref{sec:optimization} details decision-making of power grid and NetDC operators, and then introduces the bilevel optimization problem for Power-NetDC coordination. Section \ref{sec:learning} introduces the contextual regression approach for AgentCONCUR. Section \ref{sec:num_experiments} applies AgentCONCUR to New York ISO system and Section \ref{sec:conclusions} concludes.

\section{Optimizing Power--NetDC Coordination}\label{sec:optimization}
\begin{figure}
\centering
        \begin{tikzpicture}
        
        \node[solid,draw=black,thick,circle,fill=gray!10,inner sep=1pt,minimum size=10pt] (pn1) at (0,-0.15) {\scriptsize\textcolor{black}{$\boldsymbol{1}$}};
        \node[solid,draw=black,thick,circle,fill=gray!10,inner sep=1pt,minimum size=10pt] (pn2) at (1.25,0.25){\scriptsize\textcolor{black}{$\boldsymbol{2}$}};
        \node[solid,draw=black,thick,circle,fill=gray!10,inner sep=1pt,minimum size=10pt] (pn4) at (0.75,-0.85)  {\scriptsize\textcolor{black}{$\boldsymbol{...}$}};
        \node[solid,draw=black,thick,circle,fill=gray!10,inner sep=1pt,minimum size=10pt] (pn5) at (2.65,-0.6)  {\scriptsize\textcolor{black}{$\boldsymbol{...}$}};
        \node[solid,draw=black,thick,circle,fill=gray!10,inner sep=1pt,minimum size=10pt] (pn6_) at (1.85,-1.2)  {\scriptsize\textcolor{black}{$\boldsymbol{...}$}};
        \node[solid,draw=black,thick,circle,fill=gray!10,inner sep=1pt,minimum size=10pt] (pn3) at (3.25,0.45) {\scriptsize\textcolor{black}{$\boldsymbol{...}$}};
        \node[solid,draw=black,thick,circle,fill=gray!10,inner sep=1pt,minimum size=10pt] (pn6) at (5,0.3) {\scriptsize\textcolor{black}{$\boldsymbol{...}$}};
        \node[solid,draw=black,thick,circle,fill=gray!10,inner sep=1pt,minimum size=10pt] (pnb) at (6,-0.75) {\footnotesize\textcolor{black}{$\boldsymbol{b}$}};
        \node[solid,draw=black,thick,circle,fill=gray!10,inner sep=1pt,minimum size=10pt] (pn7) at (4.5,-1.25) {\footnotesize\textcolor{black}{$\boldsymbol{...}$}};
        
        \draw[draw=gray,thick] (pn1) -- (pn2);
        \draw[draw=gray,thick] (pn2) -- node[pos=0.5,sloped,fill=white] {...} (pn3);
        \draw[draw=gray,thick] (pn2) -- node[pos=0.5,sloped,fill=white] {...} (pn4);
        \draw[draw=gray,thick] (pn1) -- node[pos=0.5,sloped,fill=white] {...} (pn4);
        \draw[draw=gray,thick] (pn4) -- (pn6_);
        \draw[draw=gray,thick] (pn5) -- (pn6_);
        \draw[draw=gray,thick] (pn4) -- (pn5);
        \draw[draw=gray,thick] (pn3) -- (pnb);
        \draw[draw=gray,thick] (pn5) -- (pn7);
        \draw[draw=gray,thick] (pnb) -- (pn7);
        \draw[draw=gray,thick] (pn3) -- (pn5);
        \draw[draw=gray,thick] (pn3) -- (pn6);
        \draw[draw=gray,thick] (pn6) -- (pnb);
        \draw[draw=black,->,>=stealth,thick] (6,-0.05) node[above] {$p_{b}+r_{b}$} -- (pnb);
        \draw[draw=black,->,>=stealth,thick] (pnb) -- node[below,yshift=-0.25 cm,xshift=-0.75cm] {$d_{b}-\ell_{b}$} (5.25,-1.75);
        \draw[draw=black,->,>=stealth,thick] (pnb) -- node[pos=0.5,right] {$\gamma_{bn}\vartheta_{n}$} (6,-2);
        
        \node[solid,draw=maincolor,thick,rectangle,fill=maincolor,inner sep=1pt,minimum size=10pt] (dnn) at (6,-3.0) {\footnotesize\textcolor{white}{$\boldsymbol{n}$}};
        \node[solid,draw=maincolor,thick,rectangle,fill=maincolor,inner sep=1pt,minimum size=10pt] (dn1) at (0,-2.5) {\scriptsize\textcolor{white}{$\boldsymbol{1}$}};
        \node[solid,draw=maincolor,thick,rectangle,fill=maincolor,inner sep=1pt,minimum size=10pt] (dn2) at (1.85,-2.75) {\scriptsize\textcolor{white}{$\boldsymbol{2}$}};
        \node[solid,draw=maincolor,thick,rectangle,fill=maincolor,inner sep=1pt,minimum size=10pt] (dn3) at (3.25,-1.8) {\scriptsize\textcolor{white}{$\boldsymbol{...}$}};
    
        \draw[draw=gray,dashed,line width = 0.05cm] (pnb) -- (dnn);
        \draw[draw=gray,dashed,line width = 0.05cm] (pn3) -- (dn3);
        \draw[draw=gray,dashed,line width = 0.05cm] (pn1) -- (dn1);
        \draw[draw=gray,dashed,line width = 0.05cm] (pn6_) -- (dn2);

        \node[solid,draw=green,thick,diamond,fill=green,inner sep=1pt,minimum size=12pt] (du1) at (1,-5.25) {\scriptsize\textcolor{white}{$\boldsymbol{1}$}};
        \node[solid,draw=green,thick,diamond,fill=green,inner sep=1pt,minimum size=12pt] (du2) at (2.0,-4.6) {\scriptsize\textcolor{white}{$\boldsymbol{...}$}};
        \node[solid,draw=green,thick,diamond,fill=green,inner sep=1pt,minimum size=12pt] (dum) at (5,-4.75) {\footnotesize\textcolor{white}{$\boldsymbol{m}$}};

        \draw[green,thick,<-,>=stealth] (dn1) edge[bend right=15] (du1);
        \draw[green,thick,<-,>=stealth] (dn1) edge[bend right=15] (dum);
        \draw[green,thick,<-,>=stealth] (dn1) edge[bend right=15] (du2);
        \draw[green,thick,<-,>=stealth] (dn2) edge[bend right=15] (du1);
        \draw[green,thick,<-,>=stealth] (dn2) edge[bend right=15] (du2);
        \draw[green,thick,<-,>=stealth] (dn2) edge[bend right=15] (dum);
        \draw[green,thick,<-,>=stealth] (dn3) edge[bend right=15] (du1);
        \draw[green,thick,<-,>=stealth] (dn3) edge[bend right=15] (du2);
        \draw[green,thick,<-,>=stealth] (dn3) edge[bend left=15] (dum);
        \draw[green,thick,<-,>=stealth] (dnn) edge[bend left=15] (du1);
        \draw[green,thick,<-,>=stealth] (dnn) edge[bend left=15] (du2);
        \draw[green,thick,<-,>=stealth] (dnn) edge[bend left=15] (dum);

        \draw[maincolor,thick,->,>=stealth,dashed] (dn1) -- (dn2);
        \draw[maincolor,thick,->,>=stealth,dashed] (dn1) edge[bend left=15] (dn3);
        \draw[maincolor,thick,->,>=stealth,dashed] (dn2) edge[bend right=15] (dn3);
        \draw[maincolor,thick,<-,>=stealth,dashed] (dn2) edge[bend left=15] (dnn);
        \draw[maincolor,thick,->,>=stealth,dashed] (dn3) edge[bend left=15] (dnn);

        \draw[draw=maincolor,thick,->,>=stealth,,dashed] (dn1) to [bend right=20] node[pos=0.875,above,rotate=10] {$\varphi_{1\rightarrow n}$} (dnn);
        \draw[draw=black,->,>=stealth,thick] (4.5,-3.35) to [bend right=2.9] (dnn);
        \node[black] at (6.5,-2.75) {$\vartheta_{n}$} ;

        \draw[draw=black,<-,>=stealth,thick] (dnn) to [bend left=4.5] node[pos=0.5,right] {$w_{nm}$} (5.78,-3.75);
        \node[black] at (5,-5.25) {$\delta_{m}$};

        \begin{scope}[shift={(0,0.5)}]
        \draw[black,rounded corners=5,fill=maincolor!2] (-0.5,-6.5) -- (7.05,-6.5) -- (7.05,-8.0) -- (-0.5,-8.0) -- cycle;
        \node[draw=black,thick,circle,fill=gray!10,inner sep=1pt,minimum size=8pt, label={[label distance=0.25cm]0:{\footnotesize Power system bus}}] at (0.14,-6.75) {};
        \node[draw=maincolor,thick,rectangle,fill=maincolor,inner sep=1pt,minimum size=8pt, label={[label distance=0.25cm]0:{\footnotesize Data center}}] at (3.2,-6.75) {};
         \node[draw=green,thick,diamond,fill=green,inner sep=1pt,minimum size=9pt, label={[label distance=0.25cm]0:{\footnotesize User}}] at (5.64,-6.75) {};
        \draw[draw=gray,thick] (-0.25,-7.25) -- (0.5,-7.25) node[right] {\footnotesize Transmission line};
        \draw[draw=maincolor,thick,dashed,->,>=stealth] (-0.25,-7.75) -- (0.5,-7.75) node[right] {\footnotesize Virtual link};
        \draw[draw=gray,dashed,line width = 0.05cm] (3.25,-7.25) -- (4.,-7.25) node[right] {\footnotesize Power--NetDC coupling};
        \draw[draw=green,thick,->,>=stealth] (3.25,-7.75) -- (4.,-7.75) node[right] {\footnotesize Communication link};
        \end{scope}
        
        \end{tikzpicture}
        \caption{Interfaces and notation of the power system, network of data centers (NetDC), and communication network between data centers and users.}
        \label{fig:graphs}
\end{figure}
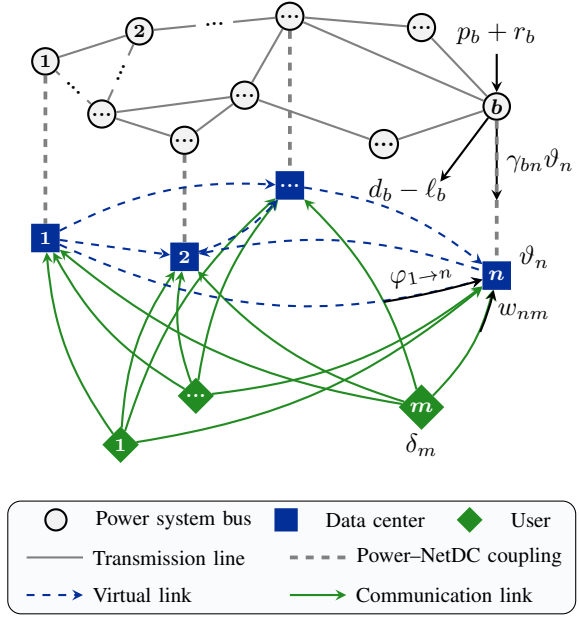

We consider the power-NetDC coordination problem, in which agents interface as illustrated in Fig. \ref{fig:graphs}. The NetDC operator selects the spatial allocation of computing tasks among data centers, receiving these tasks from a spatially distributed population of users. The main criterion for allocation is the minimization of network {\it latency} -- a time delay between sending, executing, and returning the result of a computational task for every user. The resulting task allocation subsequently shapes electricity demand, which is then incorporated into the optimal power flow (OPF) problem for power system dispatch. Thus, these two problems can be simultaneously solved in a coordinated fashion to achieve minimal dispatch costs.

The coordination is performed by means of spatial shifts of computing tasks using {\it virtual links} connecting data centers into a network \cite{zhang2020flexibility}. These shifts must be coordinated to satisfy both power system and NetDC objectives and constraints. To enable such coordination, we formulate the following bilevel optimization, where the power system operator acts as a leader, whose decision space is constrained by the NetDC operator, acting as a follower:
\begin{align*}
    \textcolor{white}
    {.}\;\;\;\!\minimize{p,\varphi}\quad&c_{\text{pwr}}(p) && \hspace{-0.5cm}{\scriptstyle\triangleright\texttt{ Dispatch cost}}\\[-1pt]
    \st\quad&p\in\mathcal{P}_{\text{pwr}}(\vartheta) && \hspace{-0.5cm}{\scriptstyle\triangleright\texttt{ Grid feasibility}}\\[2pt]
    &\minimize{\vartheta}\quad c_{\text{net-dc}}(\vartheta) &&\hspace{-0.5cm}{\scriptstyle\triangleright\texttt{ Latency loss}}\\[-1pt]
    &\st\quad\! \vartheta\in\mathcal{W}_{\text{net-dc}}(\varphi) \quad &&\hspace{-0.5cm}{\scriptstyle\triangleright \texttt{ NetDC feasibility}}
\end{align*}
Here, the task shift $\varphi$ is the {\it coordination variable}, \textcolor{blue}{which coordinates the decision-making of the two operators}. The lower-level problem takes request $\varphi$ as input, and \textcolor{blue}{finds the new task allocation $\vartheta$ among data centers that minimizes the latency loss.} The new allocation must also be feasible for the set $\mathcal{W}_{\text{net-dc}}$ of NetDC operational constraints, parameterized by $\varphi$. The optimized allocation then enters the set $\mathcal{P}_{\text{pwr}}$ of power system constraints, and the system operators computes the new generation dispatch $p$ which minimizes the dispatch cost. The optimal solution $\dot{\varphi}$ thus achieves cost-optimal and feasible for the two systems coordination. \textcolor{blue}{Importantly, this hierarchical optimization structure separates the decision-making of the two operators, and coordinates them via a shared variable $\varphi$.} 

The rest of this section details decision-making of the two systems, and then presents the coordination problem in detail.

\subsection{Power System Optimization}

\textcolor{blue}{The power system planning is based on the unit commitment (UC) problem}, which computes the least-cost \textcolor{blue}{commitment $u_{t}\in\{0,1\}^{b}$} and dispatch $p_{t}\in[\underline{p},\overline{p}]\in\mathbb{R}_{+}^{b}$ \textcolor{blue}{decisions}, that satisfies electricity net demand -- load $d_{t}\in\mathbb{R}_{+}^{b}$ subtracted by non-dispatchable renewable generation $\omega_{t}\in\mathbb{R}_{+}^{b}$. The dispatch cost is modeled using a quadratic function with the first- and second-order coefficients $c\in\mathbb{R}_{+}^{b}$ and $C\in\mathbb{S}_{+}^{b}$, respectively. The power flows are computed using the matrix of power transfer distribution factors $F\in\mathbb{R}^{b\times l}$, and must respect the line capacity $\overline{f}\in\mathbb{R}_{+}^{l}$. In rare cases, when generation lacks to satisfy all loads, we model load shedding $\ell_{t}\in\mathbb{R}^{b}$ with the most expensive cost $s \gg c$. \textcolor{blue}{The UC problem takes the form:}
\begin{subequations}\label{prob:OPF}
\begin{align}
    \minimize{p_{t},\ell_{t},\textcolor{blue}{u_{t},c_{t}^{\text{up}}}}\quad&\sum_{t=1}^{\tau}\big(c^{\top}p_{t} + p_{t}^{\top}Cp_{t} + s^{\top}\ell_{t}+\textcolor{blue}{\mathbb{1}^{\top}c_{t}^{\text{up}}}\big)\label{OPF_obj}\\
    \st\quad& \mathbb{1}^{\top}\!(p_{t}+\omega_{t}+\ell_{t}-d_{t}-\Gamma\vartheta_{t})=0,\label{OPF_bal}\\
    &|F(p_{t}+\omega_{t}+\ell_{t}-d_{t}-\Gamma\vartheta_{t})| \leqslant \overline{f},\label{OPF_flow_lim}\\
    &\textcolor{blue}{c_{t}^{\text{up}} \geqslant \mathbb{0},\quad c_{t}^{\text{up}} \geqslant c_{t}^{\text{on}}\cdot(u_{t}-u_{t-1}),}\label{OPF_su_cost}\\
    &\textcolor{blue}{\underline{p}\cdot u_{t}\leqslant p_{t}\leqslant\overline{p}\cdot u_{t},}\label{OPF_gen_one}\\
    &\textcolor{blue}{p_{t}-p_{t-1}\leqslant\overline{p}^{\uparrow}\cdot u_{t-1}} \nonumber\\
    &\quad\quad\textcolor{blue}{+ \overline{p}\cdot(1-u_{t}) +\overline{p}_{\text{up}}\cdot(u_{t}-u_{t-1}),}\\
    &\textcolor{blue}{p_{t-1}-p_{t}\leqslant\overline{p}^{\downarrow}\cdot u_{t}}\nonumber\\
    &\quad\quad\textcolor{blue}{+ \overline{p}\cdot(1-u_{t-1}) +\overline{p}_{\text{dw}}\cdot(u_{t-1}-u_{t}),}\label{OPF_gen_last}\\
    &\mathbb{0}\leqslant\ell_{t}\leqslant d_{t},\quad\textcolor{blue}{\forall t=1,\dots,\tau},\label{OPF_gen_lim}
\end{align}
\end{subequations}
which minimizes the total unit \textcolor{blue}{commitment and} dispatch cost \eqref{OPF_obj} \textcolor{blue}{across $\tau$ hours} subject to power balance condition \eqref{OPF_bal} and \textcolor{blue}{grid limits. Specifically, constraint \eqref{OPF_flow_lim} ensures that the power flows remain within the maximum transmission capacity through the scheduling horizon. The two inequalities in \eqref{OPF_su_cost} model the cost of generation unit commitment using an auxiliary variable $c_{t}^{\text{up}}$.  Generation constraints \eqref{OPF_gen_one}--\eqref{OPF_gen_last} use the binary logic to ensure that the generation units, when committed, produce within their production and ramping limits. Here, $\overline{p}^{\uparrow}$ and $\overline{p}^{\downarrow}$ are the maximum inter-hour ramping capacities, which can be different from ramping limits $\overline{p}_{\text{up}}$ and $\overline{p}_{\text{dw}}$ at a startup or shutdown of generating units, respectively.}

Modelling Power--NetDC coordination, we distinguish between conventional loads $d$ and power consumption by data centers $\Gamma\vartheta_{t}$ in constraints \eqref{OPF_bal} and \eqref{OPF_flow_lim}, where auxiliary matrix $\Gamma\in\mathbb{R}^{b\times n}$ converts computing loads $\vartheta_{t}\in\mathbb{R}^{n}$ of $n$ data centers into electrical loads. Although restrictive, this linear conversion model is consistent with power consumption models under different  utilization regimes of data centers \cite{radovanovic2021power}.

\subsection{NetDC Optimization}
The NetDC operator allocates computing tasks of $m$ users among $n$ data centers. For some computing demand $\delta_{t}\in\mathbb{R}^{m}$ \textcolor{blue}{at hour $t$}, allocation $W_{t}\in\mathbb{R}^{n\times m}$ is optimized to satisfy conservation conditions $\vartheta_{ti}=\sum_{j=1}^{m}w_{tij}$ and $\delta_{tj}=\sum_{i=1}^{n}w_{tij}$, enforced for each data center $i$ and user $j$, respectively. The goal is to minimize the latency, which is proportional to geodesic distance $G\in\mathbb{R}^{n\times m}$ between users and data centers \cite{narayanan2017right}. The proxy function for aggregated latency is defined as 
\begin{align}
\mathcal{L}:\mathbb{R}^{n\times m}\mapsto\mathbb{R}, \quad\mathcal{L}(W) = \sum_{i=1}^{n}\sum_{j=1}^{m}g_{ij}w_{ij}.\label{eq:latency}
\end{align}
The latency-optimal task allocation problem \textcolor{blue}{for each hour $t$} then becomes:
\begin{subequations}\label{prob:latency_opt}
\begin{align}
\minimize{W_{t},\vartheta_{t}\geqslant\mathbb{0}}\quad&\mathcal{L}(W_{t}) + \tfrac{\varrho}{2}\norm{W_{t}}_{2}^{2}\label{latency_opt_obj}\\
\st\quad&W_{t}^{\top}\mathbb{1}=\delta_{t},\label{latency_opt_con_1}\\
&W_{t}\mathbb{1}=\vartheta_{t},\label{latency_opt_con_2}
\end{align}
\end{subequations}
which minimizes latency subject to task conservation conditions \eqref{latency_opt_con_1} and \eqref{latency_opt_con_2}. The objective function \eqref{latency_opt_obj} additionally includes a quadratic term that evenly allocates tasks among equally remote data centers, using a small parameter $\varrho>0$.

Although the resulting data center loading $\dot{\vartheta}_{t}$ is latency-optimal, it is ignorant of the processes in the power system and may shape an expensive electricity demand allocation $\Gamma\dot{\vartheta}_{t}$. \textcolor{blue}{In this scenario, the power system may leverage the space-time load shifting flexibility of data centers to reduce the cost of grid operations. To model NetDC flexibility, consider a vector of load shifts $\varphi\in\mathbb{R}^{k}$ along $k=\big(\tfrac{n(n-1)}{2}\big)\tau + n(\tau-1)$ virtual links, connecting $n$ data centers in space and across $\tau$ time steps. Without loss of generality, we assume that tasks can be shifted to any other data center within a given time step or postponed to the adjacent time step at the same data center.} \textcolor{blue}{To model the data center connectivity, consider a space-time incidence matrix $A\in\mathbb{R}^{n\tau\times k}$ of the NetDC graph, with entries}
\begin{align}
    a_{ij} = 
    \left\{\!\!
    \begin{array}{rl}
        +1, & \text{if}\;i=n  \\
        -1, & \text{if}\;i=n' 
    \end{array}
    \right.\quad\forall j=(n,n')\in1,\dots,k,
\end{align}
\textcolor{blue}{so that the change in data center loading is computed as 
\begin{align}
A\varphi = \begin{bmatrix}\tilde{\vartheta}_{1}\\[-3pt]\dots\\\tilde{\vartheta}_{\tau}\end{bmatrix}-\begin{bmatrix}\dot{\vartheta}_{1}\\[-3pt]\dots\\\dot{\vartheta}_{\tau}\end{bmatrix},\label{virtual_links}
\end{align}
where $\tilde{\vartheta}_{t}$ is the vector of new computing loads at a particular hour $t$. Equation \eqref{virtual_links} preserves the conservation of computing loads of data centers across space and time. Advancing or postponing computing tasks of individual users results in a new task schedule $\tilde{\delta}_{1},\dots,\tilde{\delta}_{\tau}\in\mathbb{R}_{+}^{m}$, and requires enforcing the following conservation condition:
\begin{align}
    \sum_{t=1}^{\tau} \big(\tilde{\delta}_{t} - \delta_{t}\big) = 0.\label{task_conservation}
\end{align}}
\textcolor{blue}{Aswering to the load shift request $\varphi$, the NetDC operator solves the following computing task re-allocation problem:
\begin{subequations}\label{prob:SLS}
\begin{align}
    \minimize{\tilde{\vartheta}_{t},\tilde{\delta}_{t},W_{t}\geqslant\mathbb{0}}\quad&\sum_{t=1}^{\tau}\Big(\tfrac{1}{2}\norm{\mathcal{L}(W_{t}-\dot{W}_{t})}_{2}^{2} + \tfrac{\varrho}{2}\norm{\tilde{\delta}_{t}-\delta_{t}}_{2}^{2}\Big)\label{SLS_obj}\\
    \st\quad&\mathcal{L}(W_{t}-\dot{W}_{t}) \leqslant \alpha\mathcal{L}(\dot{W}_{t}),\label{SLS_balance_v4}\\
    &W_{t}^{\top}\mathbb{1}=\tilde{\delta}_{t},\quad W_{t}\mathbb{1}=\tilde{\vartheta}_{t},\quad\forall t=1,\dots,\tau,\label{SLS_balance_v1}\\
    &\text{equations}\;\eqref{virtual_links}\;\text{and}\;\eqref{task_conservation},\label{last_con}
\end{align}
\end{subequations}
where the objective function \eqref{SLS_obj} minimizes the overall latency loss as the result of altering the latency-optimal allocation $\dot{W}_{t}$, plus the regularization term which penalizes the deviation of tasks from the original schedule using a small parameter $\varrho>0$. In case the time re-allocation of tasks incurs costs to users, $\varrho$ can also be used to assign those costs. While the objective steers the solution in a latency-aware fashion, inequality \eqref{SLS_balance_v4} imposes the hard constraint on the latency loss, such that if the task shift request $\varphi_{t}$ increases the latency beyond the $\alpha\%$ of the original latency, the request is considered infeasible. Inequality constraints in \eqref{SLS_balance_v1} simply reinstate the conservation requirements for the new task schedule and data center loading, respectively. 
}

\subsection{Bilevel Optimization for Power and NetDC Coordination}
Since the vector of task shifts affects \textcolor{blue}{the UC costs in \eqref{prob:OPF} and the latency optimality loss in \eqref{prob:SLS} simultaneously, $\varphi$ is considered  a coordination variable} between power system and NetDC operators. We propose to optimize the coordination variable using the following bilevel optimization problem:
 \textcolor{blue}{\begin{align}
    &\begin{array}{@{}rl@{}}
    \minimize{p_{t},\ell_{t},u_{t},c_{t}^{\text{up}},\varphi}&\sum_{t=1}^{\tau}\big(c^{\top}p_{t} + p_{t}^{\top}Cp_{t} + s^{\top}\ell_{t}+\mathbb{1}^{\top}c_{t}^{\text{up}}\big)\!\!\!\!\!\!\!\!\\
    \st\;& \mathbb{1}^{\top}\!(p_{t}+\omega_{t}+\ell_{t}-d_{t}-\Gamma\tilde{\vartheta}_{t})=0,\\[3pt]
    &|F(p_{t}+\omega_{t}+\ell_{t}-d_{t}-\Gamma\tilde{\vartheta}_{t})| \leqslant \overline{f},\\[3pt]
    &c_{t}^{\text{up}} \geqslant \mathbb{0},\quad c_{t}^{\text{up}} \geqslant c_{t}^{\text{on}}\cdot(u_{t}-u_{t-1}),\\[3pt]
    &\underline{p}\cdot u_{t}\leqslant p_{t}\leqslant\overline{p}\cdot u_{t},\\[3pt]
    &p_{t}-p_{t-1}\leqslant\overline{p}^{\uparrow}\cdot u_{t-1} \nonumber\\[3pt]
    &\;+\; \overline{p}\cdot(1-u_{t}) +\overline{p}_{\text{up}}\cdot(u_{t}-u_{t-1}),\!\!\!\!\\[3pt]
    &p_{t-1}-p_{t}\leqslant\overline{p}^{\downarrow},\nonumber\\[3pt]
    &\;+\; \overline{p}\cdot(1-u_{t-1}) +\overline{p}_{\text{dw}}\cdot(u_{t-1}-u_{t}),\!\!\!\!\\[3pt]
    &\mathbb{0}\leqslant\ell_{t}\leqslant d_{t},\quad\forall t=1,\dots,\tau,
    \end{array}\tag{BL.U}\label{UL}
\end{align}
where the data center loading $\tilde{\vartheta}_{1},\dots,\tilde{\vartheta}_{\tau}$ comes from:
\begin{align}
    &\begin{array}{@{}rl@{}}
    \underset{\tilde{\vartheta}_{t},\tilde{\delta}_{t},W_{t}\geqslant\mathbb{0}}{\text{minimize}} & \sum_{t=1}^{\tau}\Big(\tfrac{1}{2}\norm{\mathcal{L}(W_{t}-\dot{W}_{t})}_{2}^{2} + \tfrac{\varrho}{2}\norm{\tilde{\delta}_{t}-\delta_{t}}_{2}^{2}\Big)\!\!\!\!\!\!\!\!\!\!\!\!\!\!\!\!\!\!\!\!\!\!\!\!\!\!\!\!\!\!\\[3pt]
    \st\!\!\!\!&\mathcal{L}(W_{t}-\dot{W}_{t}) \leqslant \alpha\mathcal{L}(\dot{W}_{t}),\\[3pt]
    &W_{t}^{\top}\mathbb{1}=\tilde{\delta}_{t},\quad W_{t}\mathbb{1}=\tilde{\vartheta}_{t},\\[3pt]
    &\forall t=1,\dots,\tau, \quad \text{equations}\;\eqref{virtual_links}\;\text{and}\;\eqref{task_conservation}.
    \end{array}\tag{BL.L}\label{LL}
\end{align}
In this bilevel problem, the task shift request $\varphi$ is the variable in \eqref{UL}, that parameterizes the lower-level problem \eqref{LL}, which, in turn, responses with the new data center loading. This way, the bilevel optimization identifies the cost-optimal and feasible for the two system tasks shifts.} A common solution strategy for this problem is to replace \eqref{LL} with its Karush--Kuhn--Tucker (KKT) conditions \cite{pozo2017basic}, yielding a mixed-integer problem reformulation  (see Appendix \ref{app_bilevel_ref}). 

\textcolor{blue}{While solving problem (BL) requires collecting the sensitive parameters of the two systems into one optimization problem, the privacy risks for individual NetDC users are limited, as the NetDC operator submits aggregated computing demands without disclosing any individual computational request.}

\textcolor{blue}{The bilevel problem (BL) can be solved within the day-ahead planning routines to coordinate the space-time task allocation that favors the UC costs while respecting the latency constraints of NetDC users. However, in case of deviations from the planning parameters (e.g., due to computing or electricity demand forecast errors), re-solving problem (BL) in real-time} is challenging because of large data requirements, the possible lack of real-time coordination interfaces between the power grid and NetDC operators, and the computational burden of bilevel programming that may fail to provide the solution within narrow time frames. \textcolor{blue}{Motivated by these challenges, the next section introduces the AgentCONCUR framework for efficient coordination in real-time.}


\section{Agent Coordination via Contextual Regression (AgentCONCUR) for Real Time}\label{sec:learning}
To bypass the real-time coordination challenges, we adopt a \textit{contextual} regression approach, which consists of two stages. At the first, \textcolor{blue}{offline} stage, a regression policy is trained to learn the cost-optimal tasks shifts using contextual, easy-to-access features. They may include partial yet strongly correlated with grid conditions features, such as aggregated demand, electricity prices, network flows, and renewable power generation\footnote{Importantly, such contextual information is available online from many power system operators, including the NYISO \cite{NYISOwebsite} studied in Sec. \ref{sec:num_experiments}.}. \textcolor{blue}{After being trained on many representative scenarios of gird operations, e.g., renewable power generation scenarios, the regression policy \textit{instantly} maps the unseen realization of features into task shifts at the second, operational stage.}

\textcolor{blue}{We present the AgentCONCUR framework for peak-hour coordination, which is modeled as a single period problem, in which data centers only use the spatial flexibility, considering that computational tasks may arrive on demand and can not be delayed (e.g., online streaming), yet could be dispatched elsewhere at some latency loss. In this case, we can drop index $t$ in presentation, and ignore the second term in \eqref{SLS_obj} and temporal task conservation constraint \eqref{task_conservation}. Moreover, in real time, the commitment of generation units is known and the UC problem modeled in \eqref{UL} reduces to a DC-OPF problem \cite{chatzivasileiadis2018lecture}.} Let $x$ be a vector of contextual features, and let $\phi(x)$ be the affine regression policy of the form: $$\phi(x)\triangleq\beta_{0} + \beta_{1}x\in\mathbb{R}^{k},$$ where $\beta=(\beta_{0},\beta_{1})$ are regression parameters subject to optimization. Once they are optimized, for some feature vector $\widehat{x}$, the coordination in real-time proceeds as follows: 
\begin{align}
    \varphi = 
    \left\{\begin{array}{@{}ll@{}}
    \phi(\widehat{x}), & \text{if feasible for NetDC and OPF}\\
    \mathbb{0}, & \text{otherwise}.
    \end{array}\right.\label{therule}
\end{align}
That is, implement the regression solution if the task shifts are feasible for NetDC operations and also produce an OPF-feasible electricity load profile. Otherwise, proceed with a typically more expensive yet feasible non-coordinated solution. 

\textcolor{blue}{
\begin{remark*}\normalfont
    The coordination rule in \eqref{therule} is based on DC power flow equations. In practice, the shifts under policy $\phi(\widehat{x})$ must also be verified to respect the AC power flow equations, to ensure voltage, apparent, active and reactive power feasibility. This can be achieved by solving a load flow problem, which projects the task shifts, and resulting bus injections, onto the AC power flow equations. 
\end{remark*}
}

In the remainder, we first present the base regression training, used as a benchmark approach to optimizing policy $\phi(x)$. Then, we present the proposed training optimization problem at the core of AgentCONCUR.

\subsection{Base Regression}\label{subsec:base_ML}
The base approach to optimize policy $\phi(x)$ is two-fold:
\begin{itemize}
    \item[1.]  Collect a training dataset $\{(x_{i},\dot{\varphi}_{i})\}_{i=1}^{q}$ of $q$ records, where each record $i$ includes contextual features $x_{i}$ and the optimal solution $\dot{\varphi}_{i}$ to the \textcolor{blue}{single-period} problem (BL), specific to a particular record $i$.
    \item[2.] Training the linear regression by solving
    \begin{align}
        \minimize{\norm{\beta}_{1}\leqslant\varepsilon}\quad&\frac{1}{q}\sum_{i=1}^{q}\norm{\beta_{0} + \beta_{1}x_{i} - \dot{\varphi}_{i}}_{2}^{2}
        \label{model:learning_base}
    \end{align}
    which minimizes the regularized mean squared error over $q$ historical records. Here, we chose $L_{1}-$regularization, know as {\it Lasso} \cite{tibshirani1996regression}, which encourages sparsity of $\beta$ up to selected regularization parameter $\varepsilon\in\mathbb{R}_{+}$.  For any given value $\varepsilon$, optimization \eqref{model:learning_base} selects optimal coordination features and minimizes the prediction error simultaneously.
\end{itemize} 

While being a {\it data-only} approximation of the bilevel problem solution, this approach suffers from at least two drawbacks that may prevent its practical implementation. First, although it minimizes a prediction error, it may result in large decision errors in terms of OPF costs, e.g., when under- and over-predictions of task shifts have asymmetric cost implications. This may result in a large regret, i.e., the average distance between the OPF costs induced by trained policy $\phi$ and the OPF costs of the bilevel problem (BL). Second, optimization \eqref{model:learning_base} is myopic to the upper- and lower-level feasible regions, thus risking violating operational limits of both power system and NetDC. These two observations motivate us to internalize the cost and feasibility criteria into model training. 

\subsection{Cost- and Feasibility-Aware Regression}

Optimizing policy $\phi(x)$ for AgentCONCUR, we leverage the optimization structure of bilevel model (BL) to guarantee the least-cost and feasible regression-based coordination across available historical records.  \textcolor{blue}{The proposed policy optimization for real-time coordination is based on the DC-OPF re-dispatch problem and takes the following form:}
\begin{subequations}\label{prob:bilevel_training}
\begin{align}
    \minimize{\mathcal{P}\cup\mathcal{D}\cup\varphi\cup\beta}\quad&\frac{1}{q}\sum_{i=1}^{q}\Big(\textcolor{blue}{(\dot{p}_{i}+r_{i})}^{\top}C\textcolor{blue}{(\dot{p}_{i}+r_{i})} \nonumber\\[-3pt]
    &\quad\quad\quad\quad\quad\quad\quad+c^{\top}\textcolor{blue}{(\dot{p}_{i}+r_{i})}+ s^{\top}\ell_{i}\Big)\label{bilevel_training_obj}\\
    \st\quad
    &\varphi_{i} = \beta_{0} + \beta_{1}x_{i},\;\norm{\beta}_{1}\leqslant\varepsilon,\label{bilevel_training_reg_constraint}\\
    &\mathbb{1}^{\top}\!(\textcolor{blue}{\dot{p}_{i}+r_{i}}+\omega_{i}+\ell_{i}-d_{i}-\Gamma\tilde{\vartheta}_{i})=0,\label{bilevel_training_opf_1}\\
    &|F(\textcolor{blue}{\dot{p}_{i}+r_{i}}+\omega_{i}+\ell_{i}-d_{i}-\Gamma\tilde{\vartheta}_{i})| \leqslant \overline{f},\label{bilevel_training_opf_2}\\
    &\underline{p}\leqslant \textcolor{blue}{\dot{p}_{i}+r_{i}}\leqslant\overline{p},\;\textcolor{blue}{|r_{i}|\leqslant\overline{r}},\;\mathbb{0}\leqslant\ell_{i}\leqslant d_{i},\label{bilevel_training_opf_3}\\
    &\text{KKT conditions of single-period}\;\eqref{LL},\\
    &\quad\quad\quad\quad\quad\quad\quad\quad\quad\quad\quad\forall i=1,\dots,q,\nonumber
\end{align}
\end{subequations}
\textcolor{blue}{where set $\mathcal{P}=\{(r_{i},\ell_{i},W_{i},\tilde{\vartheta}_{i})\}_{i=1}^{n}$ includes DC-OPF and NetDC primal optimization variables, and set $\mathcal{D}$ includes the dual variables of the single-period lower-level problem \eqref{LL}.} \textcolor{blue}{Relative to the base regression training in \eqref{model:learning_base}, this bilevel problem requires a larger training dataset, consisting of power system, NetDC, and contextual parameters, each specific to a particular scenario $i,$} and  minimizes the sample average OPF cost, subject to policy constraints in \eqref{bilevel_training_reg_constraint}, the set of upper-level OPF constraints in \eqref{bilevel_training_opf_1}--\eqref{bilevel_training_opf_3} and a set of KKT conditions of the lower-level problems \eqref{LL}.

\textcolor{blue}{In problem \eqref{prob:bilevel_training}, $\dot{p}_{i}$ is a vector of fixed generator dispatch from the day-ahead planning optimization, and $r_{i} \in \mathbb{R}^{b}$ is the real-time re-dispatch of generation units, which is bounded by the re-dispatch limit $\overline{r} \in \mathbb{R}^{b}$. There are two scenarios in which the coordination policy assists with real-time re-dispatch. First, when the real-time conditions deviate from the planning parameters, e.g., due to stochastic renewable generation and loads, the trained coordination policy offsets the cost incurred by re-dispatching conventional generation. Second, if the day-ahead dispatch $\dot{p}_{i}$ has been optimized in an uncoordinated fashion, the trained regression may still deliver coordination benefits in real time by re-dispatching  conventional generation within the limit $\overline{r}$. Hence, the flexibility of NetDC is limited by $\overline{r}$, which provides important trade-offs in policy optimization. A larger $\overline{r}$ allows for more flexibility and cost savings. However, larger $\overline{r}$ also withholds more reserve capacity from generation units, thus exposing grid operations to more risks. It should thus be considered as a tuning parameter of Power-NetDC coordination.}

Constraint \eqref{bilevel_training_reg_constraint} couples many instances of bilevel coordination via regression policy and its role is two-fold: it structures task shifts and selects the optimal coordination features using the $L_{1}-$regularization. \textcolor{blue}{This type of regularization also prevents overfitting and ensures robustness of the trained coordination policy in the sense of \cite{bertsimas2018characterization}.} Furthermore, this regularization bounds the optimal solution, which is necessary when $\varphi_{i} = \beta_{0} + \beta_{1}x_{i}$ is a rank-deficient system of equations, i.e., having more features than virtual links. Without this regularization, a typical optimization solver returns the status ``dual infeasible'' or ``unbounded''.

Similar to the base regression, the task shifts are restricted to the affine policy of contextual information. However, problem \eqref{prob:bilevel_training} also anticipates how the affine restriction affects the average OPF costs. Indeed, the choice of parameters $\beta$ affects the task shift requests $\varphi_{1},\dots,\varphi_{q}$, which then alter electricity load of data centers  $\vartheta_{1},\dots,\vartheta_{q}$ as they are coupled through the KKT conditions of the lower-level problem \eqref{LL}. Thus, the optimal solution of problem \eqref{prob:bilevel_training} returns regression parameters that are cost-optimal on average under the affine restriction.

Moreover, by solving problem \eqref{prob:bilevel_training}, we also guarantee the feasibility of power system and NetDC operations across historical records. Indeed, $\beta=\mathbb{0}$ is always a feasible choice, which corresponds to the latency-optimal, non-coordinated solution from problem \eqref{prob:latency_opt}. Hence, in the worst-case, problem \eqref{prob:bilevel_training} chooses a non-coordinated solution to ensure feasibility for both systems. We can also reason about the feasibility of $\phi(x)$ for unseen operational scenarios  in a probabilistic sense. Indeed, the theory of sample-based approximation of stochastic programs suggests that feasibility on unseen, out-of-sample scenarios improves as the sample size $q$ increases \cite{nemirovski2007convex,margellos2014road}. In the numerical experiments, we investigate the relationship between the sample size and the out-of-sample feasibility of the optimized coordination policy $\phi(x)$.

The training optimization \eqref{prob:bilevel_training} is solved at the operational planning stage using the similar mixed-integer reformation from Appendix \ref{app_bilevel_ref}. Although the problem is  NP-hard, modern optimization solvers, e.g., {\it Gurobi} \cite{gurobi}, make the optimization more practical than its worst-case complexity would imply. Then, at the real-time stage, the trained regression model instantly maps contextual features into computing task shifts.

\section{NYISO Case Studies}\label{sec:num_experiments}
\subsection{Data and Settings}
We use an 11-zone aggregation of the New York ISO power system depicted in Fig. \ref{fig:NYISO}, sourcing data from \cite{khan2022risk}. This zonal layout corresponds to the granularity of the contextual data from the ISO website \cite{NYISOwebsite}, which is used to train coordination policies. The power system includes approximately 30 GW of electricity demand, supplied by approximately 42 GW of conventional generation (oil, gas, and hydro) and by 1.7 GW of renewable generation (wind and solar). We install $n = 5$ data centers in the \texttt{West}, \texttt{Central}, \texttt{North}, \texttt{NYC}, and \texttt{MillWd} zones, serving customers in all $m = 11$ aggregation zones, as further shown in Fig. \ref{fig:NYISO}. \textcolor{blue}{For the day-ahead coordination, we consider that NetDC offers space-time flexibility during the evening hours from 4pm to 8pm. Computing loads can thus be shifted using $k=\big(\tfrac{n(n-1)}{2}\big)\tau + n(\tau-1)=70$ virtual links. For the peak-hour coordination by means of AgentCONCUR, we have $k = n(n-1)/2 = 10$ virtual links.} The task shifts outside the NY state area are not permitted. The computing demand $\delta_{i}$ is assumed to be proportional to the maximum peak load $d_{i}$ in the $i^{\text{th}}$ area, and will be scaled to achieve different NetDC penetration levels in the range from 5\% to 30\% of the peak system load. The operational data spans the period from January 1, 2018, to June 30, 2019, and includes 546 records, 24 hours each. All data, codes and default settings needed to replicate the results are available at 
\begin{center}
    \small\texttt{\url{https://github.com/wdvorkin/AgentCONCUR}}
\end{center}

\begin{figure}[t]
    \centering
    \includegraphics[width=0.49\textwidth]{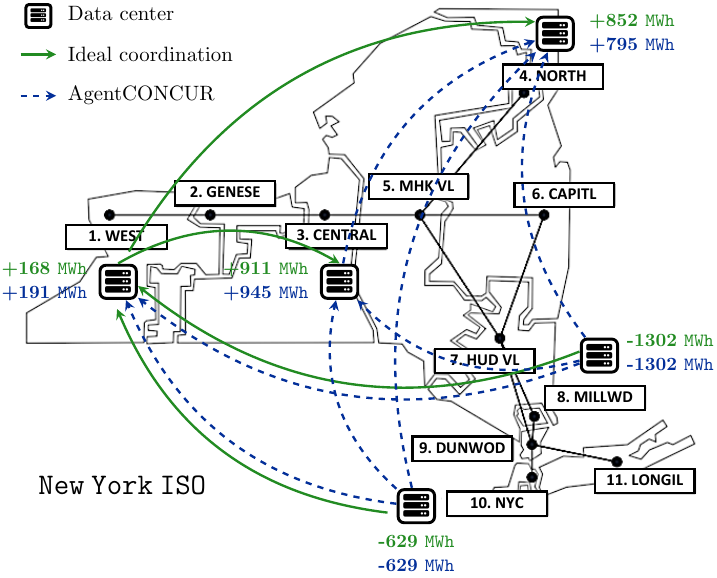}
    \caption{11-zone New York ISO system with 5 data centers. The arrows show active virtual links for peak-hour, real-time coordination under different coordination solutions for the 20\% NetDC penetration level and 100\% maximum latency loss. The change of NetDC electricity demand is given as the average across the test dataset.}
    \label{fig:NYISO}
\end{figure}

\subsection{Day-Ahead  Power-NetDC Coordination}

\textcolor{blue}{The average cost-saving potential of the day-ahead coordination is illustrated in Fig. \ref{fig:uc_cost_saving}, for the $20$\% NetDC penetration level. Here, the left plane illustrates a \$$137,812$ cost-saving potential of the temporal NetDC flexibility, when latency loss requirement is set to zero, $\alpha\rightarrow0$, thus restricting a spatial task re-allocation. The cost-saving potential then triples when invoking the spatial flexibility, by allowing the latency loss of at most $25$\% (middle plane). While with such a small latency loss the users are likely to observe no difference in the quality of service, it brings a substantial economic benefit to the power grid. Further increasing the allowable latency loss results in additional savings, yet the effect is diminishing with growing $\alpha,$ as shown in the right plane in Fig. \ref{fig:uc_cost_saving}.  This is because the additional flexibility in NetDC would result in a steep net load profile, which would require more units to commit, hence increasing the overall cost of operation. Indeed, relaxing UC constraints, i.e.,  $u\in\{1,0\} \Rightarrow 0\leqslant u \leqslant 1,$ increases the cost-saving potential from \$$445,059$ to \$$486,785$ for the latency loss of $\alpha=100\%$. This illustrates the importance of accounting for unit commitment constraints in harnessing the flexibility of data centers at the day-ahead stage.}


\textcolor{blue}{Figure \ref{fig:dc_loading} illustrates how the baseline electricity consumption of data centers changes with coordination. Relative to the baseline, the coordinated solution shifts computing tasks from the South, with the largest electricity demand center in \texttt{NYC}, to the less-loaded data centers in the Central, Western, and Northern parts of the state. At the same time, tasks are shifted in time such that the data center loading in \texttt{NYC} and \texttt{MillWd} is minimal during the peak demand hour at $6\text{pm}$.}

\textcolor{blue}{If the NetDC operator directly participates in the day-ahead market, there are implicit financial incentives to provide the grid with NetDC flexibility. Table \ref{tab:estimates} reports the average cost for NetDC electricity consumption in a non-coordinated case and three coordination cases, revealing the opportunity to reduce the cost of electricity from 6\% and 34.5\%, depending on the maximum allowable latency optimality loss.}

\begin{table}
\centering
\addtolength{\tabcolsep}{-0.4em}
\caption{\textcolor{blue}{Average charge for NetDC electricity demand during the period from 4 to 20 pm, in $\$1,000$. The coordinated solution is provided for varying latency optimality loss requirement.}}
\label{tab:estimates}
\textcolor{blue}{
\begin{tabular}{cccc}
\toprule
\multirow{2}{*}{\begin{tabular}[c]{@{}c@{}}non-coordinated \\ solution\end{tabular}} & \multicolumn{3}{c}{coordinated solution} \\
\cmidrule(lr){2-4}
 & $\alpha=0\%$ & $\alpha=25\%$ & $\alpha=100\%$ \\
\midrule
$2,804$ & $2,636\;(-6.0\%)$ & $2,328\;(-17.0\%)$  & $2,117\;(-34.5\%)$ \\
\bottomrule
\end{tabular}}
\end{table}

\textcolor{blue}{For each representative 24-hour horizon, we solve a bilevel optimization problem (BL) with $2,693$ continuous and $936$ binary variables, and  $5,611$ constraints. The latency loss $\alpha=100\%$ required longer solution time, taking  $19.9$s on average, with a standard deviation of $10.7$s, on a standard laptop.}


\begin{figure}
    \centering
    \includegraphics[width=0.48\textwidth]{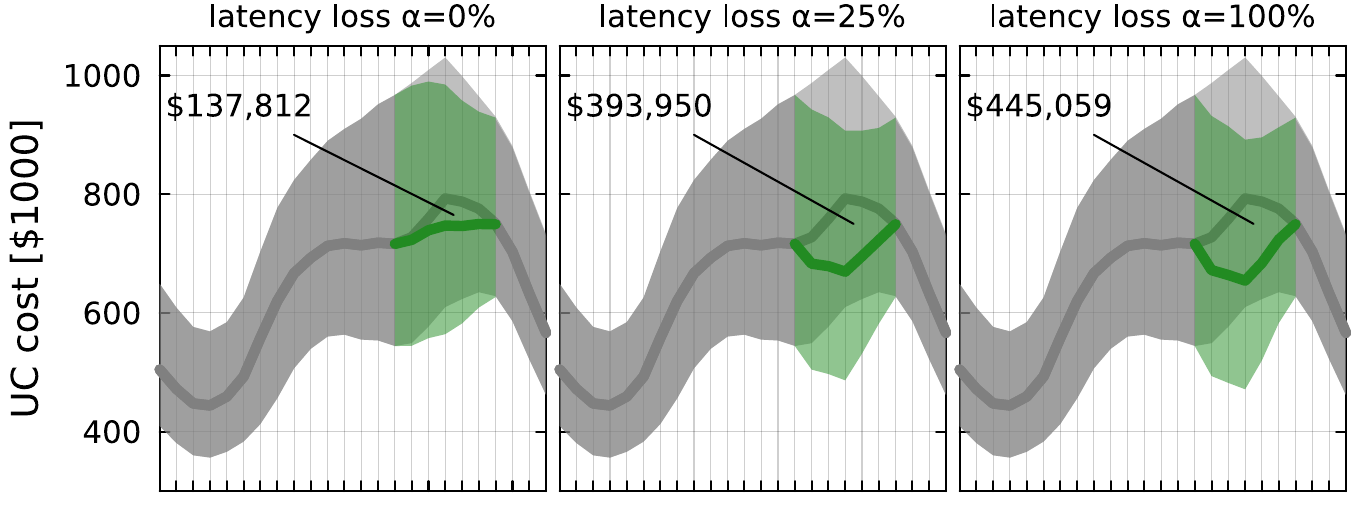}

    \caption{\textcolor{blue}{UC costs throughout the 24-hour planning horizon for three latency loss bounds. The thick lines depict the average cost and the shaded area is the confidence band. The colored area is the period of Power-NetDC coordination. The cost resulting from coordination is in green, and the non-coordinated solution is in gray. The area between the curves is the cost-saving potential.}}
    \label{fig:uc_cost_saving}
\end{figure}

\begin{figure}
    \centering
    \includegraphics[width=0.48\textwidth]{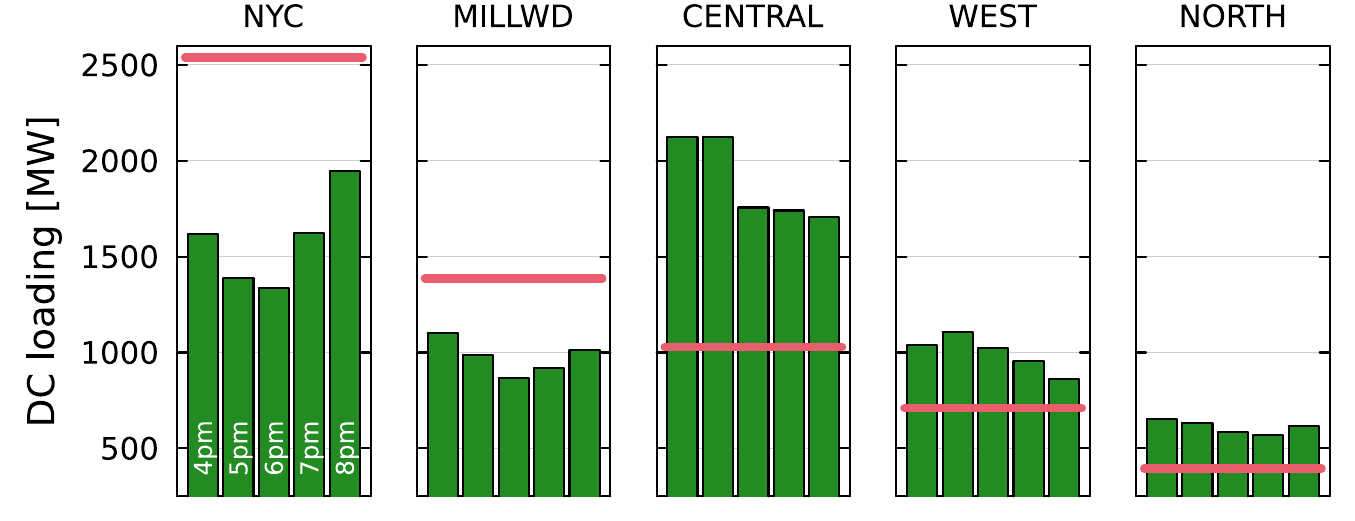}
    \caption{\textcolor{blue}{Average electricity consumption of five data centers from 4 pm to 8 pm. The horizontal lines depict the constant, baseline consumption without any coordination. The bars are hourly consumption under coordination.}}
    \label{fig:dc_loading}
\end{figure}

\subsection{Power-NetDC Coordination via AgentCONCUR}\label{subsec:agentconcur}

\textcolor{blue}{In Sec. \ref{subsec:agentconcur} through Sec. \ref{subsec:featuresel}, we consider the case of real-time peak-hour coordination by means of AgentCONCUR. We consider that the day-ahead schedule $\dot{p}$ is optimized for the nominal latency solution $\dot{\vartheta}$, and the parameter $\overline{r}$ is set to generation capacity. To train the policy, we use a dataset, where} each record contains the following contextual features, which are readily available on the New York ISO website \cite{NYISOwebsite}:
\begin{itemize}
    \item[\scriptsize\raise1.25pt\hbox{$\blacktriangleright$}] Zonal real-time electricity demand $(d)$;
    \item[\scriptsize\raise1.25pt\hbox{$\blacktriangleright$}] Zonal electricity prices $(\lambda)$;
    \item[\scriptsize\raise1.25pt\hbox{$\blacktriangleright$}] Total renewable generation, then disaggregated by zones using data on existing renewable installations $(r)$;
    \item[\scriptsize\raise1.25pt\hbox{$\blacktriangleright$}] Power flows between aggregation zones $(f)$.
\end{itemize}
Each record includes 45 contextual features, so that the coordination policy based on these features takes the form: 
\begin{align*}
    \phi
    \triangleq \beta_{0} 
    + \beta_{1}^{d}
    \begin{bmatrix}
        d_{1}\\\vdots\\d_{11}
    \end{bmatrix}
    + \beta_{1}^{\lambda}
    \begin{bmatrix}
        \lambda_{1}\\\vdots\\\lambda_{11}
    \end{bmatrix}
    + \beta_{1}^{r}
    \begin{bmatrix}
        r_{1}\\\vdots\\r_{11}
    \end{bmatrix}
    + \beta_{1}^{f}
    \begin{bmatrix}
        f_{1}\\\vdots\\f_{12}
    \end{bmatrix}
\end{align*}
To optimize and test the policy, we randomly select $q=250$ records for training and reserve the remaining 296 records for testing, unless stated otherwise. The performance of the trained coordination policies is discussed using the unseen, test set. The remaining settings include default regularization parameters $\varrho=10^{-5}$ and $\varepsilon=10$.

The dispatch costs are compared in four cases: 
\begin{itemize}
    \item[\scriptsize\raise1.25pt\hbox{$\blacktriangleright$}] {\it No coordination:} NetDC electricity demand obeys the latency-optimal solution from problem \eqref{prob:latency_opt};
    \item[\scriptsize\raise1.25pt\hbox{$\blacktriangleright$}] {\it Ideal coordination:} NetDC demand obeys the ideal coordination solution obtained from bilevel problem (BL);
    \item[\scriptsize\raise1.25pt\hbox{$\blacktriangleright$}] {\it Base regression:} NetDC demand is shifted according to the base regression policy optimized in \eqref{model:learning_base};
    \item[\scriptsize\raise1.25pt\hbox{$\blacktriangleright$}] {\it AgentCONCUR:} NetDC demand is shifted according to the regression policy optimized in \eqref{prob:bilevel_training}.
\end{itemize}

Our results reveal that the New York ISO system benefits from coordinating spatial tasks shifts in amount of $\approx 1.9$ GWh from the densely populated South towards the Central, Northern, and Western parts of the state, as shown in Fig. \ref{fig:NYISO}. Noticeably, the ideal coordination consistently uses the same 4 out of 10 virtual links, while the AgentCONCUR coordination policy enjoys more active links. This difference is due to less flexible, affine policy structure, which results in more used links to ensure feasibility across the entire training dataset simultaneously, as opposed to per-scenario feasibility satisfaction provided by the ideal coordination.

Figure \ref{fig:cost} illustrates the discrepancies in dispatch costs in all four cases. As the penetration of NetDC increases, the non-coordinated solution demonstrates rapid, quadratic growth of dispatch costs in the New York ISO dominated by conventional generation. On the other hand, the ideal coordination demonstrates a rather linear growth (e.g., see the bottom plot) of dispatch costs thanks to the cost-aware allocation of computing tasks. However, the extent of cost reduction significantly depends on the maximum allowable latency loss $\alpha,$ specified by the NetDC operator. For a small loss of 25\%, users are likely to observe no difference in the quality of service. However, this enables savings of up to 24.5\% of dispatch costs in the ideal coordination case, depending on the penetration level. The cost-saving potential increases to 49.0\% and 56.7\% in the case of double and tripled latency loss, respectively, when users experience more noticeable delays during peak-hour operations of the power system. 

This cost-saving potential is exploited by both base regression and AgentCONCUR coordination policies. However, the base regression policy, which ignores power system and NetDC operational constraints, often results in substantively higher dispatch costs, which tend to stay closer to the non-coordinated solution than to the ideal one. On the other hand, the AgentCONCUR policy, which is aware of constraints of both systems, efficiently approximates the ideal solution, i.e., staying relatively close to the ideal solution in many cases depicted in Fig. \ref{fig:cost}. However, it tends to show a larger approximation gap with the allowable latency loss and NetDC penetration increase.

\begin{figure}
    \centering
    \includegraphics[width=0.48\textwidth]{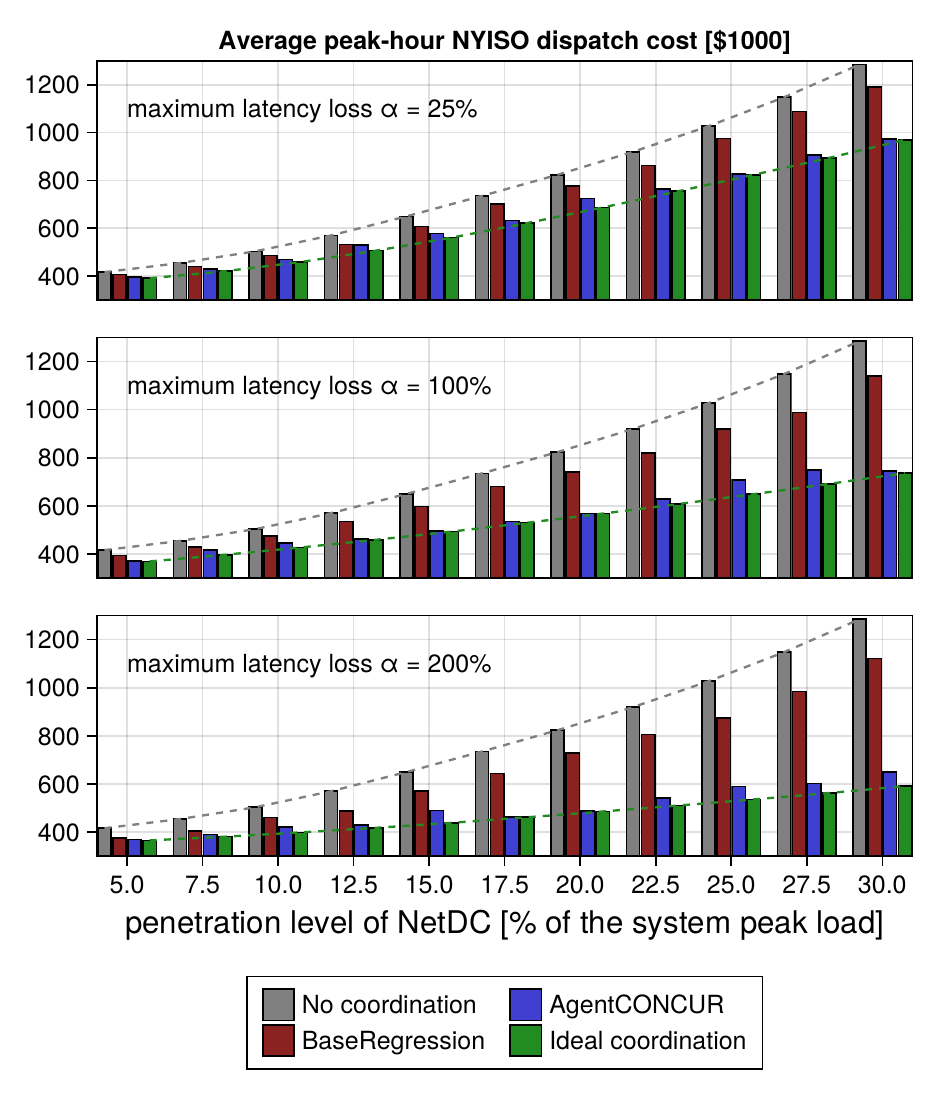}
    \caption{Average NYISO dispatch cost across the testing dataset under different coordination models for the varying NetDC penetration level and maximum allowable latency loss. The area between the dashed lines defines the cost-saving potential for regression-based coordination.}
    \label{fig:cost}
\end{figure}

\begin{table*}[b]
\caption{Selected regression features (black dots) for AgentCONCUR for different regularization parameter $\varepsilon$}
\label{tab:feature_selection}
\centering
\resizebox{1\textwidth}{!}{%
\centering
\setlength\tabcolsep{1 pt}
\begin{tabular}{rc|*{45}{c}}
\toprule
  \multicolumn{1}{c}{\multirow{2}{*}{\begin{tabular}[c]{@{}c@{}}$\varepsilon$\end{tabular}}}  
  & \multirow{2}{*}{\begin{tabular}[c]{@{}c@{}}\# of  \\ features\end{tabular}}
  &  \multicolumn{11}{c}{zonal electricity demand} & \multicolumn{12}{c}{power flow} & \multicolumn{11}{c}{zonal electricity price} & \multicolumn{11}{c}{zonal renewable power output}  \\
  \cmidrule(lr){3-13}
  \cmidrule(lr){14-25}
  \cmidrule(lr){26-36}
  \cmidrule(lr){37-47}
   & & \hspace{0.02cm} $d_{1}$ & $d_{2}$ & $d_{3}$ & $d_{4}$ & $d_{5}$ & $d_{6}$ & $d_{7}$ & $d_{8}$ & $d_{9}$ & $d_{10}$ & $d_{11}$ & $f_{1}$ & $f_{2}$ & $f_{3}$ & $f_{4}$ & $f_{5}$ & $f_{6}$ & $\lambda_{7}$ & $f_{8}$ & $f_{9}$ & $f_{10}$ & $f_{11}$ & $f_{12}$ & $\lambda_{1}$ & $\lambda_{2}$ & $\lambda_{3}$ & $\lambda_{4}$ & $\lambda_{5}$ & $\lambda_{6}$ & $\lambda_{7}$ & $\lambda_{8}$ & $\lambda_{9}$ & $\lambda_{10}$ & $\lambda_{11}$ & $r_{1}$ & $r_{2}$ & $r_{3}$ & $r_{4}$ & $r_{5}$ & $r_{6}$ & $r_{7}$ & $r_{8}$ & $r_{9}$ & $r_{10}$ & $r_{11}$ \\
   \midrule
1000.0& 29 &$\circ$& $\bullet$& $\bullet$& $\bullet$& $\circ$& $\bullet$& $\circ$& $\bullet$& $\bullet$& $\circ$& $\circ$&$\textcolor{black}{\bullet}$&$\bullet$&$\bullet$&$\bullet$&$\bullet$&$\bullet$&$\bullet$&$\bullet$&$\bullet$&$\bullet$&$\bullet$&$\circ$& 
$\bullet$& $\bullet$& $\bullet$& $\bullet$& $\bullet$& $\bullet$& $\bullet$& $\circ$& $\bullet$& $\bullet$& $\circ$& $\circ$& $\bullet$& $\bullet$& $\circ$& $\circ$& $\circ$& $\circ$& $\circ$&$\circ$&$\circ$&$\bullet$
\\
100.0& 28& $\bullet$& $\bullet$& $\bullet$& $\bullet$& $\bullet$& $\bullet$& $\circ$& $\bullet$& $\bullet$& $\bullet$& $\circ$& $\textcolor{black}{\bullet}$& $\bullet$& $\bullet$& $\bullet$& $\bullet$& $\bullet$& $\bullet$& $\bullet$& $\circ$& $\bullet$& $\circ$& $\circ$
& $\circ$& $\bullet$& $\bullet$& $\bullet$& $\bullet$& $\circ$& $\circ$& $\bullet$& $\bullet$& $\bullet$& $\bullet$& $\circ$& $\bullet$& $\circ$& $\circ$& $\circ$& $\circ$& $\circ$& $\circ$& $\circ$& $\circ$& $\bullet$
\\
10.0& 24& $\bullet$& $\bullet$& $\bullet$& $\bullet$& $\circ$& $\bullet$& $\bullet$& $\bullet$& $\bullet$& $\bullet$& $\bullet$& $\textcolor{black}{\bullet}$& $\bullet$& $\bullet$& $\bullet$& $\circ$& $\bullet$& $\bullet$& $\bullet$& $\bullet$& $\bullet$& $\circ$& $\circ$& $\circ$& $\circ$& $\circ$&$\circ$& $\circ$& $\circ$& $\circ$& $\bullet$& $\bullet$& $\bullet$& $\circ$& $\circ$& $\bullet$& $\circ$& $\circ$& $\circ$& $\circ$& $\circ$& $\circ$& $\circ$& $\circ$& $\bullet$
\\
5.0& 20& $\bullet$& $\bullet$& $\circ$& $\bullet$& $\bullet$& $\bullet$& $\bullet$&$\bullet$& $\bullet$& $\bullet$& $\bullet$& $\textcolor{black}{\bullet}$& $\bullet$& $\bullet$& $\circ$& $\bullet$& $\circ$& $\bullet$& $\bullet$& $\bullet$& $\circ$& $\circ$& $\circ$& $\circ$& $\circ$& $\circ$& $\circ$& $\circ$& $\circ$& $\circ$& $\circ$& $\bullet$& $\bullet$& $\circ$& $\circ$& $\bullet$& $\circ$& $\circ$& $\circ$& $\circ$& $\circ$& $\circ$& $\circ$& $\circ$& $\circ$
\\
2.5& 13& $\bullet$& $\circ$& $\bullet$& $\circ$& $\circ$& $\bullet$& $\bullet$& $\bullet$& $\bullet$& $\bullet$& $\circ$& $\bullet$& $\bullet$& $\circ$& $\circ$& $\circ$& $\bullet$& $\circ$& $\circ$& $\bullet$& $\circ$& $\circ$& $\circ$& $\circ$& $\circ$& $\circ$& $\circ$& $\circ$& $\circ$& $\circ$& $\circ$& $\bullet$& $\bullet$& $\circ$& $\circ$& $\circ$& $\circ$& $\circ$& $\circ$& $\circ$& $\circ$& $\circ$& $\circ$& $\circ$& $\circ$
\\
1.0& 6& $\bullet$     & $\circ$& $\bullet$& $\circ$& $\circ$& $\circ$& $\circ$& $\circ$& $\circ$& $\bullet$& $\circ$& $\textcolor{black}{\bullet}$& $\bullet$& $\circ$& $\circ$& $\circ$& $\circ$& $\circ$& $\circ$& $\circ$& $\bullet$& $\circ$& $\circ$& $\circ$& $\circ$& $\circ$& $\circ$& $\circ$& $\circ$& $\circ$& $\circ$& $\circ$& $\circ$& $\circ$& $\circ$& $\circ$& $\circ$& $\circ$& $\circ$& $\circ$& $\circ$& $\circ$& $\circ$& $\circ$& $\circ$
\\
0.5& 3& $\bullet$& $\circ$& $\circ$& $\circ$& $\circ$& $\circ$& $\circ$& $\circ$& $\circ$&$\bullet$& $\circ$& $\textcolor{black}{\circ}$& $\circ$& $\bullet$& $\circ$& $\circ$& $\circ$& $\circ$& $\circ$& $\circ$& $\circ$& $\circ$& $\circ$& $\circ$& $\circ$& $\circ$& $\circ$& $\circ$& $\circ$& $\circ$& $\circ$& $\circ$& $\circ$& $\circ$& $\circ$& $\circ$& $\circ$& $\circ$& $\circ$& $\circ$& $\circ$& $\circ$& $\circ$& $\circ$& $\circ$
\\
0.1& 1& $\circ$
& $\circ$& $\circ$& $\circ$& $\circ$& $\circ$& $\circ$& $\circ$& $\circ$& $\bullet$& $\circ$& $\textcolor{black}{\circ}$& $\circ$& $\circ$& $\circ$& $\circ$& $\circ$& $\circ$& $\circ$& $\circ$& $\circ$& $\circ$& $\circ$& $\circ$& $\circ$& $\circ$& $\circ$& $\circ$& $\circ$& $\circ$& $\circ$& $\circ$& $\circ$& $\circ$& $\circ$& $\circ$& $\circ$& $\circ$& $\circ$& $\circ$& $\circ$& $\circ$& $\circ$& $\circ$& $\circ$
\\
\bottomrule
\end{tabular}%
}
\end{table*}

\subsection{Feasibility of Regression-Based Coordination}

The approximation gaps reported in Fig. \ref{fig:cost} are due to infeasible task shifts, i.e., the shifts that violate power system constraints, NetDC constraints, or both. Whenever the task shift is infeasible in real-time, the two operators resort to a more expensive yet feasible non-coordinated solution. However, the feasibility of regression-based coordination improves with a larger size of the training dataset, as illustrated in Fig. \ref{fig:feas}. The AgentCONCUR policy dominates the base one and achieves zero violations of power system constraints (e.g., no load shedding) with sample size $q\geqslant150$. Moreover, for $q\geqslant150$, it keeps infeasibility of NetDC operations below 7\%. The dominance of AgentCONCUR is {\it consistent}, which is important when the size of the training dataset with representative records is limited. We also observed similar results across other NetDC penetration and latency parameters.

\subsection{CPU Times}
While increasing the size of the training dataset improves feasibility, it also increases the computational burden of problem \eqref{prob:bilevel_training}, as further shown in Fig. \ref{fig:CPU_time}. For a reasonable choice of $q=150$, the CPU times are approximately $8$ hours. However, since the regression training by means of problem \eqref{prob:bilevel_training} is the off-line process which takes place at the operational planning stage, the CPU time is considered reasonable. In real-time, at the time of coordination, the mapping of contextual features into coordination actions is instant.

\begin{figure}
    \centering
    \includegraphics[width=0.48\textwidth]{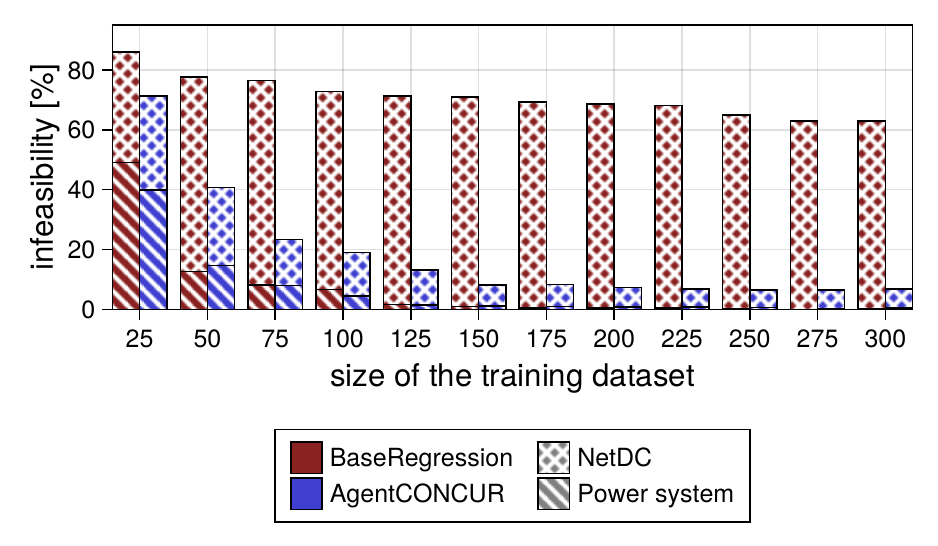}
    \caption{Infeasibility of regression-based coordination as the function of the training dataset size. Results are for 20\% NetDC penetration and 25\% maximum latency loss, averaged across 100 random draws of training scenarios.}
    \label{fig:feas}
\end{figure}

\begin{figure}
    \centering
    \includegraphics[width=0.48\textwidth]{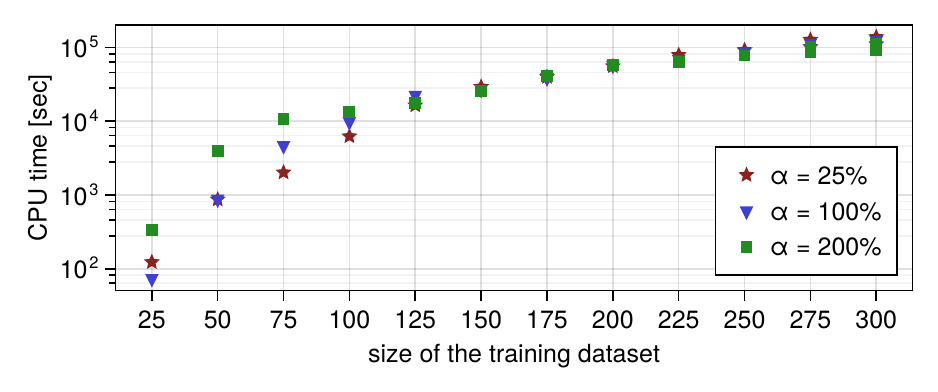}
    \caption{Average CPU times to solve the mixed-integer reformulation of the bilevel program \eqref{prob:bilevel_training}. NetDC penetration level is 20\%.}
    \label{fig:CPU_time}
\end{figure}

\subsection{Coordination Feature Selection}\label{subsec:featuresel}

\begin{figure}
    \centering
    \includegraphics[width=0.48\textwidth]{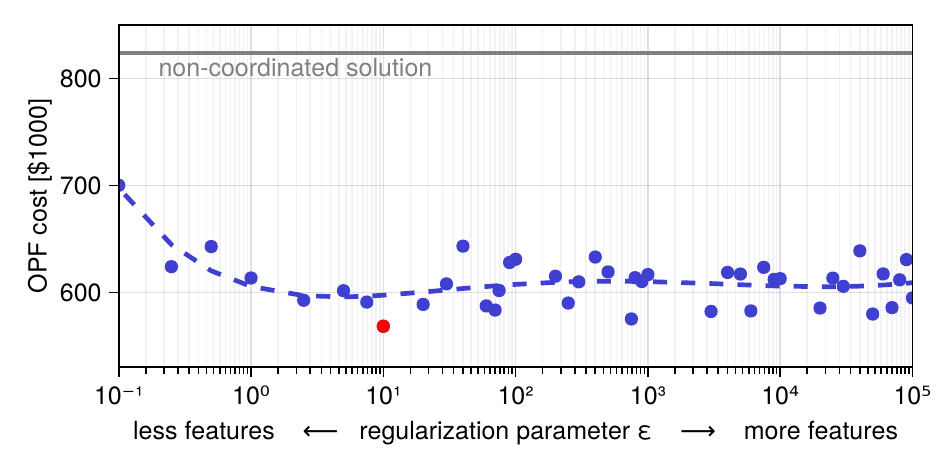}
    \caption{OPF costs for varying regularization parameter $\varepsilon$. The blue dots depict the average costs obtained on the test dataset, and the dash line is the trend. 
    The red dot marks the optimal selection that minimizes the cost on average.}
    \label{fig:feature_selection}
\end{figure}

While 45 contextual features are used in training, we demonstrate that the Power--NetDC coordination can also be achieved with fewer features, i.e., with less data requirements. 

We perform the feature selection by tuning regularization parameter $\varepsilon$ in the regression policy training problem \eqref{prob:bilevel_training}. The smaller the $\varepsilon$, the fewer features are used by the policy. Table \ref{tab:feature_selection} reports the selected features for various assignments of $\varepsilon$. Observe, as the feature space shrinks ($\downarrow \varepsilon$), the policy gives less priority to renewable power output, which is reasonable as the NYISO has a small installed renewable capacity at present (e.g., only 1.7 GW). As the space further shrinks, less priority is given to electricity prices, that indeed become less informative in uncongested cases. The power flows and electricity demands, on the other hand, consistently present among selected features for AgentCONCUR.

Figure \ref{fig:feature_selection} further reveals the trade-off between the dispatch cost and amount of selected features. Approximately, the same level of costs (see the dashed trend line) can be achieved in the range $\varepsilon\in[1,10^5]$, selecting from 6 to $30+$ features. Moreover, parameter $\varepsilon$ can be optimized to achieve the optimal dispatch cost under regression-based coordination. Here, the optimal $\varepsilon^\star=10$  selects 24 features for coordination.

Notably, for $\varepsilon=0.1$, the coordination of task shifts within the entire NetDC is performed with a {\it single} feature, i.e., electricity demand of the largest demand center -- New York City. Although this is not the cost-optimal choice, this is the least data-intensive coordination, which still performs better than the non-coordinated solution, as also shown in Fig. \ref{fig:feature_selection}.

\section{Conclusions}\label{sec:conclusions}

To streamline the economic coordination of power grids and data centers, this work proposed to transition from data-intensive optimization-based coordination to a light weighted regression-based coordination. Recognizing the risks of trusting a regression model with coordinating two critical infrastructure systems, we devised a new training algorithm which inherits the structure of the optimization-based coordination and enables feasible and cost-consistent computing task shifts in real-time. The case study on NYISO system with various NetDC penetration levels revealed 24.5--56.7\% cost-saving potential, most of which has shown to be delivered by regression policies at different data-intensity preferences.   

There are some notable limitations that motivate several directions for future work. For example, while the optimization-based coordination remunerates data center flexibility via duality theory \cite{zhang2022remunerating}, as such, the duality-based solution is unavailable under regression policies. It is thus relevant to study the integration of regression policies into real-time electricity markets. Furthermore, although the proposed mechanism does not require any private data exchange at the time of coordination, it still needs sensitive data from the power system and NetDC for training at the planning stage. One potential solution to remove this practical bottleneck is the use of \textcolor{blue}{differentially private algorithms for data obfuscation} \cite{dvorkin2023differentially}, yet it will require additional modifications to the training procedure to eliminate the effect of noise.

\section*{Acknowledgements}
This work was in part supported by the Marie Skłodowska-Curie Actions COFUND Postdoctoral Program, Grant Agreement No. 101034297 -- project Learning ORDER.

\appendix
\subsection{Mixed-Integer Reformulation of the Bilevel Problem}\label{app_bilevel_ref}

\begin{subequations}\label{eq:KKTs}
The Karush--Kuhn--Tucker conditions of the lower-level problem \eqref{LL} are derived from the following Lagrangian:\textcolor{blue}{\begin{align*}
      &\underset{\mu}{\text{max}}\underset{W,\vartheta,\tilde{\delta}}{\text{min}}\;\; \sum_{t=1}^{\tau}\tfrac{1}{2}\norm{\mathcal{L}(W_{t})-\mathcal{L}(\dot{W}_{t})}_{2}^{2} 
      + \sum_{t=1}^{\tau}\tfrac{\varrho}{2}\norm{\tilde{\delta}_{t}-\delta_{t}}_{2}^{2} \\
      & -\sum_{t=1}^{\tau}\mu_{\delta t}^{\top}(W_{t}^{\top}\mathbb{1} - \tilde{\delta}_{t}) 
        -\sum_{t=1}^{\tau}\mu_{\vartheta t}^{\top}(W_{t}\mathbb{1} - \tilde{\vartheta}_{t})\\
      & -\sum_{t=1}^{\tau}\langle\mu_{w t},W_{t}\rangle_{\text{F}}  -\sum_{t=1}^{\tau}\mu_{\alpha t} (-\mathcal{L}(W_{t}-\dot{W}_{t}) + \alpha\mathcal{L}(\dot{W}_{t})) \\
      & -\mu_{\varphi}^{\top}\left(A\varphi - \begin{bmatrix}\tilde{\vartheta}_{1}\\[-3pt]\dots\\\tilde{\vartheta}_{\tau}\end{bmatrix}+\begin{bmatrix}\dot{\vartheta}_{1}\\[-3pt]\dots\\\dot{\vartheta}_{\tau}\end{bmatrix}\right) -\mu_{\tau}\Big(\sum_{t=1}^{\tau}(\tilde{\delta}_{t}-\delta_{t})\Big)
\end{align*}
}
The stationarity conditions are as follows:
\begin{align}
    \tilde{\vartheta}_{t}&\colon\mu_{\vartheta t} + \mu_{\varphi} = \mathbb{0},\label{eq:KKT_stat_1}\\
    \textcolor{blue}{\tilde{\delta}_{t}}&\textcolor{blue}{\colon\varrho(\tilde{\delta}_{t}-\delta_{t})+\mu_{\delta t} -\mu_{\tau} = \mathbb{0},}\label{eq:KKT_stat_2}\\
     w_{tij}&\colon g_{ij}(\mathcal{L}(W_{t}-\dot{W}_{t})) - \mu_{\delta t j} \!- \mu_{\vartheta t i} - \mu_{w t ij} \! + \mu_{\alpha t} g_{ij} = 0, \nonumber\\
     &\quad\quad\quad\quad\!\!\!\!\!\!\!\! \forall t=1,\dots,\tau,\; i=1,\dots,n,\;j=1,\dots,m.\label{eq:KKT_stat_3}
\end{align}
The dual feasibility conditions are:
\begin{align}\label{eq:KKT_dual_feas}
&\mu_{w t ij}\geqslant0,\quad\forall i=1,\dots,n,\;j=1,\dots,m,\\
&\mu_{\alpha t}\geqslant0,\quad\forall t=1,\dots,\tau.
\end{align}
The non-convex complementarity slackness conditions are addressed using an equivalent SOS1-based reformulation \cite{siddiqui2013sos1}:
\begin{align}
    &\{\mu_{w t ij},w_{tij}\}\in\text{SOS1},\;\forall i=1,\dots,n,\;j=1,\dots,m,\\
    &\{\mu_{\alpha t},\mathcal{L}(W_{t}\!-\!\dot{W}_{t})\!-\!\alpha\mathcal{L}(\dot{W}_{t})\}\!\in\!\text{SOS1},\forall t=1,\dots,\tau.
\end{align}
where notation $\{x,y\}\in\text{SOS1}$ means that at most one variable may be nonzero. The equivalent reformulation of problem (BL) is then obtained when the lower-level problem \eqref{LL} is replaced with constraints \eqref{SLS_balance_v4}--\eqref{last_con} and \eqref{eq:KKTs}.
\end{subequations}

\balance
\bibliography{references.bib}
\bibliographystyle{IEEEtran}

\endgroup
 \end{document}